\documentclass[10pt]{article}

\begin{document}

\def\CA{{\cal A}}
\def\CB{{\cal B}}
\def\CC{{\cal C}}
\def\CD{{\cal D}}
\def\CE{{\cal E}}
\def\CF{{\cal F}}
\def\CG{{\cal G}}
\def\CH{{\cal H}}
\def\CI{{\cal I}}
\def\CJ{{\cal J}}
\def\CK{{\cal K}}
\def\CL{{\cal L}}
\def\CM{{\cal M}}
\def\CN{{\cal N}}
\def\CO{{\cal O}}
\def\CP{{\cal P}}
\def\CQ{{\cal Q}}
\def\CR{{\cal R}}
\def\CS{{\cal S}}
\def\CT{{\cal T}}
\def\CU{{\cal U}}
\def\CV{{\cal V}}
\def\CW{{\cal W}}
\def\CX{{\cal X}}
\def\CY{{\cal Y}}
\def\CZ{{\cal Z}}

\newcommand{\todo}[1]{{\em \small {#1}}\marginpar{$\Longleftarrow$}}
\newcommand{\labell}[1]{\label{#1}\qquad_{#1}} 
\newcommand{\bbibitem}[1]{\bibitem{#1}\marginpar{#1}}
\newcommand{\llabel}[1]{\label{#1}\marginpar{#1}}

\newcommand{\sphere}[0]{{\rm S}^3}
\newcommand{\su}[0]{{\rm SU(2)}}
\newcommand{\so}[0]{{\rm SO(4)}}
\newcommand{\bK}[0]{{\bf K}}
\newcommand{\bL}[0]{{\bf L}}
\newcommand{\bR}[0]{{\bf R}}
\newcommand{\tK}[0]{\tilde{K}}
\newcommand{\tL}[0]{\bar{L}}
\newcommand{\tR}[0]{\tilde{R}}

\newcommand{\ack}[1]{[{\bf Ack!: {#1}}]}

\newcommand{\btzm}[0]{BTZ$_{\rm M}$}
\newcommand{\ads}[1]{{\rm AdS}_{#1}}
\newcommand{\ds}[1]{{\rm dS}_{#1}}
\newcommand{\dS}[1]{{\rm dS}_{#1}}
\newcommand{\eds}[1]{{\rm EdS}_{#1}}
\newcommand{\sph}[1]{{\rm S}^{#1}}
\newcommand{\gn}[0]{G_N}
\newcommand{\SL}[0]{{\rm SL}(2,R)}
\newcommand{\cosm}[0]{R}
\newcommand{\hdim}[0]{\bar{h}}
\newcommand{\bw}[0]{\bar{w}}
\newcommand{\bz}[0]{\bar{z}}
\newcommand{\be}{\begin{equation}}
\newcommand{\ee}{\end{equation}}
\newcommand{\bea}{\begin{eqnarray}}
\newcommand{\eea}{\end{eqnarray}}
\newcommand{\pat}{\partial}
\newcommand{\lp}{\lambda_+}
\newcommand{\bx}{ {\bf x}}
\newcommand{\bk}{{\bf k}}
\newcommand{\bb}{{\bf b}}
\newcommand{\BB}{{\bf B}}
\newcommand{\tp}{\tilde{\phi}}
\hyphenation{Min-kow-ski}

\newcommand{\pa}{\partial}
\newcommand{\eref}[1]{(\ref{#1})}

\def\apr{\alpha'}
\def\str{{str}}
\def\lstr{\ell_\str}
\def\gstr{g_\str}
\def\Mstr{M_\str}
\def\lpl{\ell_{pl}}
\def\Mpl{M_{pl}}
\def\varep{\varepsilon}
\def\del{\nabla}
\def\grad{\nabla}
\def\tr{\hbox{tr}}
\def\perp{\bot}
\def\half{\frac{1}{2}}
\def\p{\partial}
\def\perp{\bot}
\def\eps{\epsilon}

\newcommand{\BC}{\mathbb{C}}
\newcommand{\BR}{\mathbb{R}}
\newcommand{\BZ}{\mathbb{Z}}
\newcommand{\bra}[1]{\langle{#1}|}
\newcommand{\ket}[1]{|{#1}\rangle}
\newcommand{\vev}[1]{\langle{#1}\rangle}
\newcommand{\Real}{\mathfrak{Re}}
\newcommand{\Imag}{\mathfrak{Im}}
\newcommand{\talpha}{{\widetilde{\alpha}}}
\newcommand{\Ham}{{\widehat{H}}}
\newcommand{\al}{\alpha}
\newcommand\x{{\bf x}}
\newcommand\y{{\bf y}}

\def\NPB{{\it Nucl. Phys. }{\bf B}}
\def\PL{{\it Phys. Lett. }}
\def\PRL{{\it Phys. Rev. Lett. }}
\def\PRD{{\it Phys. Rev. }{\bf D}}
\def\CQG{{\it Class. Quantum Grav. }}
\def\JMP{{\it J. Math. Phys. }}
\def\SJNP{{\it Sov. J. Nucl. Phys. }}
\def\SPJ{{\it Sov. Phys. J. }}
\def\JETPL{{\it JETP Lett. }}
\def\TMP{{\it Theor. Math. Phys. }}
\def\IJMPA{{\it Int. J. Mod. Phys. }{\bf A}}
\def\MPL{{\it Mod. Phys. Lett. }}
\def\CMP{{\it Commun. Math. Phys. }}
\def\AP{{\it Ann. Phys. }}
\def\PR{{\it Phys. Rep. }}

\renewcommand{\thepage}{\arabic{page}}
\setcounter{page}{1}

\rightline{hep-th/0507253}
\rightline{VPI-IPPAP-05-02}

\vskip 0.75 cm
\renewcommand{\thefootnote}{\fnsymbol{footnote}}
\centerline{\Large \bf Quantum Gravity, Torsion, Parity Violation and all that}
\vskip 0.75 cm

\centerline{{\bf
Laurent Freidel,${}^{1,2}$\footnote{lfreidel@perimeterinstitute.ca}
Djordje Minic${}^{3}$\footnote{dminic@vt.edu} and
Tatsu Takeuchi${}^{3}$\footnote{takeuchi@vt.edu}}}
\vskip .5cm
\centerline{${}^1$\it Perimeter Institute for Theoretical Physics,}
\centerline{\it 31 Caroline St. N. Waterloo, N2L 2Y5, ON, Canada. }
\vskip .5cm
\centerline{${}^2$\it Laboratoire de Physique, Ecole Normale Sup{\'e}rieure de Lyon}
\centerline{\it 46 All{\'e}e d'Italie, 69364 Lyon, Cedex 07, France.}
\vskip .5cm
\centerline{${}^3$\it Institute for Particle Physics and Astrophysics,}
\centerline{\it Department of Physics, Virginia Tech}
\centerline{\it Blacksburg, VA 24061, U.S.A.}
\vskip .5cm

\setcounter{footnote}{0}
\renewcommand{\thefootnote}{\arabic{footnote}}

\begin{abstract}
We discuss the issue of parity violation in quantum gravity. In particular,
we study the coupling of fermionic degrees of freedom in the presence of
torsion and the physical meaning of the Immirzi parameter from the viewpoint of effective
field theory.
We derive the low-energy effective lagrangian which turns out to involve {\it two} parameters, one measuring the non-minimal coupling
of fermions in the presence of torsion, the other being the Immirzi parameter. In the case
of non-minimal coupling the effective lagrangian contains an axial-vector interaction leading to
parity violation.
Alternatively, in the case of minimal coupling there is no parity violation and the effective lagrangian
contains only the usual axial-axial interaction. In this situation the real values of the
Immirzi parameter are not at all constrained. 
On the other hand, 
purely imaginary values of the Immirzi parameter lead to violations of unitarity for the case of non-minimal coupling.
Finally, the effective lagrangian blows up for the positive and negative unit imaginary values of the Immirzi parameter.
\end{abstract}

\newpage

\section{Quantum Gravity and Effective field theory}
\noindent
The Wilsonian point of view on effective field theory is a very powerful organizing principle
allowing us to drastically restrict the number of theories relevant for the description of
low energy physics.
As far as the low energy physics is concerned, the Wilsonian approach instructs us to consider the most general local actions
invariant under a given symmetry, while forgetting about the underlying microscopic degrees of freedom.
Of course this philosophy might have to be radically changed in the presence of gravity, but in this
article we adopt the usual point of view in order to understand better the issue of gravitationally induced parity violation.

Gravity is a theory which is classically invariant under the diffeomorphism group and since we
want to be able to incorporate Standard model-like fermions, the effective field theory should also be invariant under a local Lorentz
gauge symmetry. The dynamical field variables are given by the frame field
(a one form valued in the vectorial representation)
$e_\mu^Idx^\mu$ and Lorentz connection (a one form valued in the Lorentz Lie algebra)
$\omega_\mu^{IJ}dx^\mu$.
If we apply the effective field theory point of view to gravity, and consider only actions
which are analytic functionals of the frame field connection and its derivatives\footnote{This excludes
functionals which depend on the inverse frame field.}, we easily read out the most relevant terms in the low energy
effective action. To leading order this low energy effective action contains exactly $6$ possible terms.
Three of these terms are topological invariants analogous to the QCD theta term. They correspond to:
the Euler class
\begin{equation}
\theta_1\int R^{IJ}(\omega) \wedge R^{KL}(\omega)\epsilon_{IJKL},
\end{equation}
the Pontryagin class
\begin{equation}
\theta_2 \int R^{IJ}(\omega) \wedge R_{IJ},
\end{equation}
and the Nieh-Yan class
\begin{equation}
\theta_3 \int d_\omega e^I \wedge d_\omega e_I - R^{IJ}\wedge e_I\wedge e_{J},
\end{equation}
where $R^{IJ}(\omega)= d\omega^{IJ} + \omega^{IK}\wedge\omega_K{}^J$ is the canonical curvature two form.
Since we are interested in semi-classical physics we concentrate our attention to the bulk terms\footnote{Topological
``deformations'' of Einstein's classical theory in four dimensions have been investigated in detail in \cite{jackiw}.}.
Two of the bulk terms are well known; they correspond to the Einstein-Hilbert action
and the cosmological constant term. The corresponding coupling constants are, respectively, the Newton constant
$G$ and the cosmological constant $\Lambda$.
This is, however, not the final answer, for there exists an additional term which is usually 
disregarded. From an
effective field theory point of view, this term is as important as the two others, and thus has to be
taken into account in the low energy effective action. The corresponding
coupling constant is called the Immirzi parameter, $\gamma$.

Therefore, the leading bulk terms in the most general four-dimensional low energy gravitational action, in any microscopic quantum theory of
gravity (be it string theory, loop gravity etc.) are given by
\begin{equation}
\widetilde{S}_P = -\frac{1}{32\pi G}\int
\left(  R^{IJ}(\omega)\wedge e^K \wedge e^L \epsilon_{IJKL} -
\frac{\Lambda}{6} e^I \wedge e^J \wedge e^K \wedge e^L \epsilon_{IJKL}
-\frac{2}{\gamma}R^{IJ}(\omega)\wedge e_I \wedge e_J
\right) \label{actionfinal}
\end{equation}
Note that unlike the Newton and cosmological constant,
the Immirzi parameter is dimensionless.
This action and the meaning of the Immirzi parameter was first discussed by Holst \cite{Holst}.
Usually $\gamma$ is taken to be equal to zero; 
this forces the torsion to be the null\footnotemark
\footnotetext{The action (\ref{actionfinal}) is singular when
$\gamma =0$. One can however introduce an auxiliary field $B^{I}$ (a two form
valued in the vectorial representation) in order to rewrite the Immirzi term, up to a boundary contribution,
as
$\int B^{I}\wedge T_{I} + \gamma T^I\wedge T_I$, where $T^I=d_\omega e^I$ is the torsion.
When $\gamma=0$ the equation of motion for the $B$ field forces the torsion to be null.}
and we recover the usual second order formulation of gravity.
It is also frequent but less common to encounter the choice $\gamma=\infty$. In this case
we recover the Cartan-Weyl formulation of gravity \cite{torsion}.

From the effective field theory point of view there is a priori no reason to prefer one choice or another, and thus,
one should take into account all possible values of $\gamma$ and let experiment decide its numerical
value. This is the point of view which is taken here. Effectively, at the quantum level, this amounts to treating
$\gamma$ as a superselection parameter.
We should add a caveat here: as briefly mentioned above,
in this paper we assume the usual local Wilsonian effective theory, in which decoupling
between the UV and IR degrees of freedom is usually taken for granted. This nevertheless does not have
to be the case in the presence of gravity \cite{uvir}, but we refrain from any further discussion of this important issue 
in this letter \cite{future}.

Classically when there is no matter coupled to gravity the equations of motion for the connection imply zero
torsion irrespective of the value of the Immirzi parameter.
At the quantum level this is not the case and one expects observable effects associated with a 
particular microscopic formulation of quantum gravity (such as loop quantum gravity) which 
contains this parameter \cite{laurent}, \cite{carlo}.
In \cite{laurent} it is shown that $\gamma$ controls the rate of fluctuation of the torsion at the 
quantum level and when different from $0$ or $\infty$, it leads 
to a compactification of the phase space of gravity.
The Immirzi parameter plays an important role in loop quantum gravity, which predicts, for example, that the area of a surface or
the volume of a spatial region are quantized \cite{Ashtekar} (the unit quanta of area and volume being $\gamma l_P^2$ and
$\gamma^{3/2} l_P^3$). The spectra of the area and volume operators are discrete provided that
the Immirzi parameter is not equal to zero.
In contrast, there is no indication that area or volume are quantized in string theory 
(which might be because of the
the subtle issue of background independence).
Thus any direct experimental constraint of the Immirzi parameter can be taken as concrete steps towards the falsification of
a specific microscopic approach. This is the main motivation for the analysis that follows.

\section{Torsion and Fermions}
In this section we follow the derivation of Rovelli and Perez \cite{carlo},
obtaining similar but crucially different results. We start with
the formulation of general relativity in the first order formalism
in terms of a Lorentz connection $\omega_{\mu}{}^{IJ}$ and a frame
field $e_\mu^I$, where $I,J\ldots = 0,1,2,3$ denote the
internal Lorentz indices and $\mu,\nu \ldots = 0, 1,2,3$  the respective
space-time indices. The curvature is defined to be
$R_{\mu\nu}^{IJ}(\omega) = \partial_\mu \omega_\nu^{IJ}-
\partial_\nu \omega_\mu^{IJ} + [\omega_\mu,\omega_\nu]^{IJ}$.
When the frame field is invertible
the
gravitational action (\ref{actionfinal}) which includes the Immirzi
parameter and a zero cosmological constant can be written in the
form
\begin{equation}
 S_{G}[e,\omega]=\frac{1}{16\pi G}\int d^4x\ e e^{\mu}_{I}
e^{\nu}_{J}\ P^{IJ}{}_{KL} R^{KL}_{\mu \nu}(\omega), \label{actionf2}
\end{equation}
where we have introduced the following tensor and its inverse
\begin{equation}
P^{IJ}{}_{KL}=\delta^{[I}_{K}\delta^{J]}_{L}
-\frac{1}{\gamma}\frac{\epsilon^{IJ}{}_{KL}}{2}, \quad \quad
P^{-1}{}_{KL}{}^{IJ}=\frac{\gamma^{2}}{\gamma^2+1}\left(\delta^{[I}_{K}\delta^{J]}_{L}
+\frac{1}{\gamma}\frac{\epsilon^{IJ}{}_{KL}}{2}\right).
\end{equation}
The
coupling of gravity to fermions is given by the action
\begin{equation}
S_{F}[e,\omega,\psi]= \int d^4x\ \frac{ie}{2}
\left(\overline\psi\, \gamma^I e_{I}^\mu \nabla_{\mu} \psi
-\overline{\nabla_{\mu}\psi}\, \gamma^I e_{I}^\mu  \psi\right),
\label{actionf}
\end{equation}
where $\gamma^I$ are the Dirac
matrices\footnotemark \footnotetext{We use the particle physicist
conventions $\{\gamma^I,\gamma^J\}=2\eta^{IJ}$ with
$\eta^{IJ}=\mathrm{diag}(+1,-1,-1,-1)$, $\gamma_{5}=\gamma^{5}= i
\gamma^{0}\gamma^{1}\gamma^{2}\gamma^{3}$. The reality conditions
are $\gamma_{I}^\dagger= \gamma_{0}\gamma_{I}\gamma_{0}$,
$(i\gamma_{5})^\dagger= \gamma_{0}(i\gamma_{5})\gamma_{0}$},
and
\begin{equation}
\nabla_{\mu}\equiv \partial_{\mu}+ \omega_{\mu}^{IJ}
\frac{\gamma_{[I}\gamma_{J]}}{4},\quad \quad
[\nabla_{\mu},\nabla_{\nu}]=
R_{\mu\nu}^{IJ}\frac{\gamma_{[I}\gamma_{J]}}{4}.
\end{equation}
We can
decompose $\omega_{\mu}^{IJ}$ as
$\omega_{\mu}^{IJ}=\widetilde{\omega}_{\mu}^{IJ}+ C_{\mu}^{IJ}$,
where $\widetilde{\omega}$ is the torsion free spin connection
satisfying
\begin{equation}
\widetilde{\nabla}_{[\mu}e^I_{\nu]}=0
\end{equation}
and
$C_{\mu}{}^{IJ}$ is the so called `contorsion' tensor related to
the torsion in the obvious way
${\nabla}_{[\mu}e_{\nu]}{}^I=C_{[\mu}{}^{IJ}e_{\nu] J}$.

It is important to consider in the fermion action the above real
combination (a point overlooked in \cite{carlo}) in order to get
the same equation of motion for $\psi$ and $\overline{\psi}$.
The general fermion action written in terms of the covariant derivative $\nabla$,
which gives back the usual fermion
action in the absence  of gravity, can be written in the form 
\begin{equation}
\tilde{S}_{F}[e,\omega,\psi]=
\int d^4x\ \frac{ie}{2}
\left((1-i\alpha) \overline\psi\,\gamma^I e_{I}^\mu \nabla_{\mu} \psi
-(1+i\alpha) \overline{\nabla_{\mu}\psi}\, \gamma^I e_{I}^\mu  \psi\right).
\label{actionff}
\end{equation}
Such an action {\it does not} give the same equation of motion
for  $\psi$ and $\overline{\psi}$ unless
\begin{equation}
(\alpha -\alpha^*) e_{I}^\mu C_{\mu}{}^{IJ}=0.
\end{equation}
If $\alpha$ is not real this imposes a constraint on the torsion tensor
which is incompatible with the gravitational equation of motion that fixes the
torsion.
This is due to the fact that the following symmetric combination
is {\it{not}} a total derivative in the presence of torsion \cite{torsion}
\be \label{symm}
\frac{e}{2}\left(\overline\psi\, \gamma^I e_{I}^\mu \nabla_{\mu} \psi
+\overline{\nabla_{\mu}\psi}\, \gamma^I e_{I}^\mu  \psi\right)
=
\partial_{\mu}(e e_{I}^\mu V^I) + e_{I}^\mu C_{\mu}{}^{I}{}_{J} V^J
\ee
where $V^I$ is the vector fermion current ($A^I$ denotes the
axial fermion current)
\begin{equation}
V^I = \overline\psi\, \gamma^I \psi, \quad \quad A^I = \overline\psi\,
\gamma^{5}\gamma^I \psi,
\end{equation}
Both $V^I$ and $A^I$ are real currents.
The value $\alpha=0$ corresponds to the usual minimal coupling of fermions to gravity.
In general, an arbitrary real value for $\alpha$ corresponds to a non-minimal coupling.

We now consider the equation of motion coming from the variation of equation
(\ref{actionf}) with respect to the
connection $\delta{ (S_G +S_F)}/\delta \omega= 0$.
Since
\begin{equation}
\frac{\delta S_G}{\delta \omega_\mu{}^{IJ}} =
\frac{ 1}{8\pi G} \nabla_\nu( e e^\mu_Ke^\nu_L)P^{KL}{}_{IJ}
\end{equation}
and
\begin{equation}
\frac{\delta S_F}{\delta \omega_\mu{}^{IJ}} =
\frac{ie e_K^{\mu}}{8} \overline{\psi} \{\gamma_K, \gamma_{[I} \gamma_{J]}\} \psi,
\end{equation}
this equation of motion reads\footnotemark \footnotetext{We use the identities
$\overline{\Gamma \psi} = \overline{\psi} \gamma_0 \Gamma^\dagger \gamma_0$,
$\{\gamma_K, \gamma_{[I} \gamma_{J]}\} = -2i \epsilon_{IJKL}\gamma^5 \gamma^L$,
${[}\gamma_K, \gamma_{[I} \gamma_{J]}{]}= 4 \eta_{K[I}\gamma_{J]}$}
\begin{equation}
\frac{1}{2\pi G}\left(C_{ [KL]}{}^{\mu} + C_{\nu [K}{}^{\nu} e_{L]}{}^\mu\right)P^{KL}{}_{IJ}
=  {\epsilon_{IJ}{}^{KL}} e^{\mu}_K A_L.
\end{equation}
The solution is given by
\begin{equation}
e^\mu_{I} C_{\mu J K}= 4 \pi G \frac{\gamma^2}{\gamma^{2}+1}
\left(\frac{1}{2}\epsilon_{IJKL}A^{L} -\frac{1}{\gamma} \eta_{I[J}A_{K]}\right).
\end{equation}
If one does the same computation starting with the action (\ref{actionff})
$\tilde{S}_{F}$
one obtains instead
\begin{equation}
e^\mu_{I} C_{\mu J K}= 4 \pi G \frac{\gamma^2}{\gamma^{2}+1}
\left(\frac{1}{2}\epsilon_{IJKL}(A+\frac{\alpha}{\gamma}V)^{L}
-\frac{1}{\gamma} \eta_{I[J}(A-\alpha \gamma V)_{K]}\right).
\end{equation}
It is clear from this expression that
$e^\mu_{I} C_{\mu}{}^{IK} \propto \gamma/(\gamma^{2}+1) (A-\alpha \gamma V)^{K}$ is different
from $0$ unless $\gamma =0 $ or $\gamma=\infty$\footnote{\label{RL}Another theoretical possibility is to have the constraint $A^K=\alpha \gamma V^K$ satisfied. This is possible 
if we have $\alpha\gamma =\pm 1$. In this case this constraint is satisfied if all fermions are right handed ($\alpha\gamma=-1$)
 or left handed ($\alpha\gamma=-1$). However this theoretical possibility is clearly excluded by experiments, since we see both right and left handed fermions and we 
 will therefore not discuss this option in what follows.
 Note that this argument relies on the fact that in our framework the non minimal coupling $\alpha$ is supposed to be the same for all species.
 It is logically possible to relax this constraint and study the case of a ``species-dependent'' $\alpha$ \cite{Andrew}. We will not study this possibility which seems unnatural.}. 
 This shows, as previously stated, that one should take $\alpha$ real in order
to have a consistent action principle.

We can now obtain an equivalent action by replacing $\omega$ with
$\widetilde{\omega}+C(\psi)$ in (\ref{actionf2}) and (\ref{actionff}). 
The terms of (\ref{actionf2}) linear in the
fermion current are total derivatives, leaving us with the
standard second order tetrad action of general relativity with fermions,
plus a four fermi interaction term 
$S_{int}[e,\psi]=S_{int}^{(1)}[e,\psi]+S_{int}^{(2)}[e,\psi]$.
This interaction term comes first from the evaluation of
\begin{equation}
S_{int}^{(1)}[e,\psi]= \frac{1}{16 \pi G} \int d^4x\ e e^{\mu}_{I}
e^{\nu}_{J}\ P^{IJ}{}_{KL} [C_\mu, C_\nu]^{KL}.
\end{equation}
One can use the following identity
\begin{equation}
e^{\mu}_{I}
e^{\nu}_{J}\ P^{IJ}{}_{KL} [C_\mu, C_\nu]^{KL}
= 6\left( U^2 -\frac{2UV}{\gamma} - V^2\right)
\end{equation}
if $e^\mu_{I} C_{\mu J K}=\epsilon_{IJKL}U^{L} +  \eta_{IJ}V_{K}-\eta_{IK}V_{J}$,
in order to get  
\begin{equation}
S_{int}^{(1)}[e,\psi]=
\frac{3}{2}\pi G \left(\frac{\gamma^2}{\gamma^2+1}\right) \left( A^2 + \frac{2\alpha}{\gamma}A\cdot V
-\alpha^2 V^2\right).
\end{equation}
The other contribution is given by
\be
S_{int}^{(2)}[e,\psi]= \int d^4x\ \frac{e}{4} e^\mu_{I}C^{IJK}
\left(\epsilon_{IJKL}A^{L} + 2 \alpha \eta_{I[J}V_{K]}\right)
\ee
leading to the final expression for the effective action
\begin{equation}
S_{int}[e,\psi]=
-\frac{3}{2}\pi G \left(\frac{\gamma^2}{\gamma^2+1}\right) \left( A^2 + \frac{2\alpha}{\gamma}A\cdot V
-\alpha^2 V^2\right).
\end{equation}
Note that this effective action is similar yet different in detail from the one derived in
\cite{carlo}.
One sees that the only interaction which is independent of $\alpha$ is the axial-axial
interaction. The parity violating vector-axial interaction is absent if one considers the minimal
coupling $\alpha=0$. Thus parity violation {\it cannot} be taken as a measure of
the Immirzi parameter but only of the combination $\frac{\alpha \gamma}{\gamma^2 +1}$ depending on the non-minimal coupling parameter.
Note also that when the Immirzi parameter is purely imaginary $\gamma=\pm i$, the interaction becomes infinite. This infinite factor multiplies 
the term $(A\mp i\alpha V)^2$ and effectively imposes the constraint between vector and axial-vector currents discussed in footnote \ref{RL}.

We now turn to the discussion of experimental significance of the gravitational
parity violation for both non-minimal and minimal couplings of fermions.

\section{Parity Violation: The Weak Charge}
 
In this section we review the necessary physics needed to discuss possible gravitational
parity violation.

We start with the low energy effective Lagrangian describing interactions between electrons and quarks
induced by $Z$ exchange
\begin{equation}
{\cal L}_{eff}
= -\frac{8G_F}{\sqrt{2}}\rho \sum_q
   \Bigl[ g_L^q\bar{q}_L\gamma^\mu q_L
+ g_R^q\bar{q}_R\gamma^\mu q_R
   \Bigr]
   \Bigl[ g_L^e\bar{e}_L\gamma_\mu e_L
+ g_R^e\bar{e}_R\gamma_\mu e_R
   \Bigr]\;,
\label{Lequark}
\end{equation}
where 
\begin{eqnarray}
g_L^f & = & I_f - Q_f\sin^2\theta_W \;, \cr
g_R^f & = & \phantom{I_f} - Q_f\sin^2\theta_W \;.
\end{eqnarray}
$I_f$ and $Q_{f}$ are respectively, the isospin and charge of the fermion.
The parity violating part of Eq.~(\ref{Lequark}) is
\begin{equation}
{\cal L}_{PV}
= \frac{G_F}{\sqrt{2}} \sum_q
  \Bigl[ C_{1q}(\bar{q}\gamma^\mu q)(\bar{e}\gamma_\mu \gamma_5 e)
       + C_{2q}(\bar{q}\gamma^\mu \gamma_5 q)(\bar{e}\gamma_\mu e)               
  \Bigr]\;,
\label{LPV}
\end{equation}
where
\begin{eqnarray}
C_{1q} &=& 2\rho( g_L^q + g_R^q )( g_L^e - g_R^e )\;,\cr
C_{2q} &=& 2\rho( g_L^q - g_R^q )( g_L^e + g_R^e )\;.
\label{ConeCtwo}
\end{eqnarray}
The first term of Eq.~(\ref{LPV}) leads to a parity violating interaction
between electrons and nuclei whose amplitude is given by
\begin{equation}
\mathcal{M} = \frac{G_F}{\sqrt{2}}
\sum_q C_{1q} \!\int\!\!d^4x
\bra{ Z, N } \bar{q}(x)\gamma^\mu q(x)
\ket{ Z, N }
\bra{ e^-(p_f) } \bar{e}(x)\gamma_\mu \gamma_5 e(x)
\ket{ e^-(p_i) }\;,
\label{AmpZN}
\end{equation}
Where $\ket{Z,N}$ denotes a nucleus consisting of
$Z$ protons and $N$ neutrons.
 
In the non--relativistic or static
limit of the nucleus, we can neglect the
space components of the quark vector current
matrix element.
Its time component
\begin{equation}
  \bra{Z,N} \bar{q}(x)\gamma^0 q(x)\ket{Z,N}
= \bra{Z,N} q^\dagger(x)       q(x)\ket{Z,N}
= \rho_q({\bf r})
\label{qdensity}
\end{equation}
yields the density of quark flavor $q$ inside the nucleus.
Since the size of the nucleus can be assumed to be small
compared to
the wavelength of the momentum transfer $(p_f-p_i)$,
we can make the approximation
\begin{equation}
\rho_q(\mathbf{r}) \approx N_q \,\delta^{(3)}(\mathbf{r})
\label{monopole}
\end{equation}
where $N_q$ is the total number of quarks of flavor $q$
contained in the nucleus.
Therefore,
\begin{eqnarray}
\sum_q C_{1q} \bra{Z,N} \bar{q}(x)\gamma^0 q(x) \ket{Z,N}
&=& \left[ C_{1u} (2Z+N) + C_{1d} (Z+2N)
   \right] \delta^3({\bf r}) \cr
&\equiv& -\frac{1}{2}Q_W(Z,N)\, \delta^{(3)}({\bf r})\;,
\label{uddensity}
\end{eqnarray}
where $(g_L^e-g_R^e)=-\frac{1}{2}$ 
in $C_{1q}$ is factored out so that
$Q_W(Z,N)$ depends only on the nucleus.
 
Next, using the
relation between Dirac spinors $u(p)$ (in the Dirac representation)
and non-relativistic Pauli spinors $\phi$,
\begin{equation}
u(p) = \sqrt{E+m}
       \left[ \begin{array}{c} \phi \\
\frac{\vec{\sigma}\cdot\vec{p}}{E+m}\,\phi
\end{array}
       \right],
\label{DiracPauli}
\end{equation}
we find that the time component of
the electron axial--vector current matrix element
in the non--relativistic limit
becomes
\begin{eqnarray}
 \bra{ e^-(p_f) } \bar{e}(x)\gamma_0 \gamma_5e(x)
 \ket{ e^-(p_i) }
&=& \frac{1}{\sqrt{2E_fV}} \frac{1}{\sqrt{2E_iV}} \;
   \bar{u}(p_f)\gamma_0 \gamma_5 u(p_i)\; e^{-i(p_i-p_f)x} \cr
&=& \frac{1}{2m_eV}\;
   \phi_f^\dagger
   \left[ {\vec{\sigma} \cdot \vec{p}_f}
+ {\vec{\sigma} \cdot \vec{p}_i}
   \right]
   \phi_i \,e^{i(\mathbf{p}_i - \mathbf{p}_f)\cdot \mathbf{r}}\;.
\label{NRapprox}
\end{eqnarray}
 
Inserting Eq.~(\ref{uddensity}) and (\ref{NRapprox}) into (\ref{AmpZN}),
we find
\begin{equation}
\mathcal{M}  = -\frac{G_F}{4\sqrt{2}m_e}\, Q_W(Z,N)
\int\!\! d^4x\;
\varphi_f^\dagger(x)
\left[  \hat{\vec{\sigma}}\cdot\hat{\vec{p}}
\;\delta^3(\hat{\mathbf{r}})
+ \delta^3(\hat{\mathbf{r}})\;
\hat{\vec{\sigma}}\cdot\hat{\vec{p}}
\right]
\varphi_i(x)\;,
\label{AmpZNtwo}
\end{equation}
where
\begin{equation}
\varphi(x) = \frac{1}{\sqrt{V}}\;e^{-ipx}\phi\;.
\label{WaveFunction}
\end{equation}
From Eq.~(\ref{AmpZNtwo}), we conclude that
the first term of Eq.~(\ref{LPV})
induces a parity violating potential of the form
\begin{equation}
\hat{V}_{PV} = \frac{G_F}{4\sqrt{2}m_e}\, Q_W(Z,N)
    \left[ \hat{\vec{\sigma}}\cdot\hat{\vec{p}}
\;\delta^3( {\bf \hat{r}} )
+ \delta^3( {\bf \hat{r}} )\;
\hat{\vec{\sigma}}\cdot\hat{\vec{p}}
    \right]
\label{VPV}
\end{equation}
where $m_e$, $\hat{\vec{\sigma}}/2$, $\hat{\vec{p}}$, and
$\hat{\vec{r}}$
are respectively the mass, spin, momentum, and position
of the electron.
The factor
$Q_W(Z,N)$ is called the `weak charge' of the nucleus
and can be large enough for heavy nuclei to make the effects
of this potential observable.

The second term of Eq.~(\ref{LPV}) induces a potential
dependent on the nuclear spin \cite{Bouchiat}
which is too weak to be observed.

\section{Quantum Gravity and Parity Violation}

\subsection{Non-minimal coupling}

According to the above discussion (section 2) the effective lagrangian contains
a parity violating term provided the coupling of fermions in the presence of torsion is
non-minimal, i.e. $\alpha \neq 0$. 

\begin{equation}
\mathcal{L}_\mathrm{PV}
= \frac{3}{2} \pi G_N \frac{2 \alpha \gamma}{\gamma^2+1}\,e\,
(\bar{\psi}\gamma_A\psi)(\bar{\psi}\gamma^A\gamma_5\psi)\;,
\end{equation}
If we allow $\gamma$ to be purely imaginary
the overall effective coupling in the
low energy lagrangian is imaginary ($\alpha$ being real), and thus unitarity is
violated! Thus the purely imaginary values of the Immirzi parameter are excluded
by appealing to unitarity. Furthermore,
the contact interaction
blows up at $\gamma=\pm i$. Note that $\gamma = i$ corresponds to the self-dual Ashtekar canonical
formalism \cite{carlo}. 

Let $\gamma$ be real and let
\begin{equation}
\frac{2\gamma}{\gamma^2+1} \alpha \equiv \beta\;,\qquad
-\infty<\beta<\infty\;.
\end{equation}
Then the above effective interaction becomes
\begin{equation}
\mathcal{L}_\mathrm{PV}
= \frac{3}{2}\pi \beta\, G_N
(\bar{\psi}\gamma_\mu\psi)(\bar{\psi}\gamma^\mu\gamma_5\psi)\;,
\end{equation}
which has the same form as the first term of Eq.~(\ref{LPV}) if we assign
the vector current to the quarks and the axial-vector current to the
electrons:
\begin{equation}
\mathcal{L}_\mathrm{PV}
= \frac{3}{2}\pi \beta\, G_N
(\bar{q}\gamma_\mu q)(\bar{e}\gamma^\mu\gamma_5 e)\;,
\end{equation}
Therefore, the parity-violating interaction amplitude is
\begin{equation}
\mathcal{M}_\mathrm{PV} = \frac{3}{2}\pi \beta\,G_N
\sum_q \!\int\!\!d^4x
\bra{ Z, N } \bar{q}(x)\gamma^\mu q(x)
\ket{ Z, N }
\bra{ e^-(p_f) } \bar{e}(x)\gamma_\mu \gamma_5 e(x)
\ket{ e^-(p_i) }\;.
\end{equation}
Using
\begin{equation}
  \bra{Z,N} \bar{q}(x)\gamma^0 q(x)\ket{Z,N}
= \bra{Z,N} q^\dagger(x)       q(x)\ket{Z,N}
= \rho_q(\mathbf{r})
\approx N_q \,\delta^{(3)}(\mathbf{r})\;,
\end{equation}
we have
\begin{eqnarray}
\sum \bra{Z,N} \bar{q}(x)\gamma^0 q(x) \ket{Z,N}
= 3(Z+N)\,\delta^{(3)}(\mathrm{r})\;,
\end{eqnarray}
and using Eq.~(\ref{NRapprox}) we find
\begin{equation}
\mathcal{M}_\mathrm{PV}
  = \frac{9\pi \beta\,G_N(Z+N)}{4m_e}\,
\int\!\! d^4x\;
\varphi_f^\dagger(x)
\left[  \hat{\vec{\sigma}}\cdot\hat{\vec{p}}
\;\delta^3(\hat{\mathbf{r}})
+ \delta^3(\hat{\mathbf{r}})\;
\hat{\vec{\sigma}}\cdot\hat{\vec{p}}
\right]
\varphi_i(x)\;,
\end{equation}
Comparison with the $Z$-exchange case shows that the
corresponding coefficients are
\begin{equation}
-G_F\,Q_W(Z,N) \quad\leftrightarrow\quad
9\pi \sqrt{2}\beta\, G_N (Z+N)\;,
\end{equation}
so that the `effective' weak charge for the Immirzi parameter is
\begin{equation}
Q_I = -9\pi \beta(Z+N)\left(\frac{\sqrt{2}G_N}{G_F}\right)\;.
\end{equation}
The values of $G_N$ and $G_F$ are \cite{pdbook}
\begin{eqnarray}
G_N & = & 6.707(10)\times 10^{-39}\,(\hbar c) (\mathbf{GeV}/c^2)^{-2} \;,\cr
G_F & = & 1.16639(1)\times 10^{-5}\,(\hbar c)^3(\mathbf{GeV})^{-2} \;.
\end{eqnarray}
So in natural units, the ratio of $\sqrt{2} G_N$ to $G_F$ is about $10^{-33}$.
The factor $9\pi$ is about $3 \times 10^1$, and the factor $(Z+N)$ is about
$10^2$ for heavy nuclei, so the effective weak charge is
about $10^{-30}\beta$. 

The weak charges of heavy nuclei are typically of order $10^2$, and the
experimental errors on them are of order $1$ \cite{atomicp}.  So the experimental constraint on
$\beta$ will typically be
\begin{equation}
\beta \equiv \frac{2\gamma}{\gamma^2+1} \alpha < \mbox{about $10^{30}$}
\end{equation}
which is practically no constraint at all.
Obviously the physical reason for this is the weakness of gravity as compared to
weak interactions, and the fact that parity is already maximally violated in the weak sector.

The crucial point here is that the parity violating interaction in principle
contains {\it two} undetermined parameters,
the strength of the non-minimal coupling of fermions to gravity, and the Immirzi parameter.
This seems to have been overlooked in the literature.

\subsection{Minimal coupling}

In this situation ($\alpha=0$) the effective lagrangian contains only the axial-axial coupling and the experimental bound on the effective
coupling is known in the literature both on four dimensional \cite{torsion} and large extra dimensional
physics \cite{shapiro}, \cite{laynam}.
The effective action reads
\begin{equation}
\mathcal{L}_\mathrm{AA}
= -\frac{3}{2} \pi G_N \frac{ \gamma^{2}}{\gamma^2+1}\,e\,
(\bar{\psi}\gamma_A\gamma_5\psi)(\bar{\psi}\gamma^A\gamma_5\psi)\;,
\end{equation}
where we set $\frac{3}{2} \pi G_N \frac{ \gamma^{2}}{\gamma^2+1}\equiv \frac{3\pi}{\Lambda_T^2}$
in order to compare with \cite{shapiro}, \cite{laynam}.

For example, the axial-axial contact four-fermi interaction can affect the electron-quark contact interactions.
A typical bound is \cite{shapiro}, \cite{laynam}
\begin{equation}
\Lambda_T \ge 5.3 \,\mathrm{TeV}
\end{equation}
A stronger constraint is implied by astrophysical data provided one assumes the existence of light sterile neutrinos.
>From supernova data one infers \cite{laynam}
\begin{equation}
\Lambda_T \ge 210 \,\mathrm{TeV}
\end{equation}
If we assume that $\gamma$ is real, then
\begin{equation}
\frac{\gamma^{2}}{\gamma^2+1} < 1\;.
\end{equation}
Thus there is no bound on $\gamma$ in this case, given the usual value of the
Planck scale.

\section{Conclusions}
To summarize:
In this letter we have discussed the issue of parity violation in quantum gravity.
In particular, we have clarified the role of the coupling of fermionic degrees of freedom in the presence of
torsion as well as the physical meaning of the Immirzi parameter in the low-energy effective lagrangian.
The low energy effective theory is found to contain two parameters: the
non-minimal coupling parameter ($\alpha$) and the Immirzi parameter ($\gamma$). Only in the case
of non-minimal coupling ($\alpha \neq 0$) the effective lagrangian contains the axial-vector interaction leading to
parity violation. The important point here is that the parity violating vector-axial interaction contains {\it two} parameters ($\alpha$ and $\gamma$).
In this situation, the
experimental constraint of an effective parameter involving both $\alpha$ and $\gamma$ can be discussed. 
In the case of minimal coupling ($\alpha=0$) there is no parity violation and the effective lagrangian
contains only the usual axial-axial interaction. Here the bounds on the effective coupling, a function of
the Immirzi parameter, are well known in the literature. These do not impose any constraint on the real value of $\gamma$.  On the other hand, 
purely imaginary values of the Immirzi parameter lead to violations of unitarity for $\alpha \neq 0$.
Finally, the effective lagrangian blows up for $\gamma = \pm i$.

{\bf Acknowledgments:}
It is our pleasure to thank A. Perez and C. Rovelli for discussions of their work and A. Randono for a correspondence on the
issue of a purely imaginary $\gamma$. We also thank R. Jackiw for drawing our attention to his work with S.-Y. Pi.
The research of 
{\small DM} and  {\small TT} is supported in part by the U.S. Department
of Energy under contract DE-FG05-92ER40677.

\end{document}